\documentclass[preprint,number]{elsarticle}
\usepackage{graphicx}
\usepackage{dcolumn}
\usepackage{bm}
\usepackage{longtable}
\usepackage{color}
\usepackage{bbm}
\usepackage{amssymb,amsmath}
\usepackage{hyperref}

\newcommand{\ee}{\mathrm{e}}							
\newcommand{\im}{\mathrm{i}}	
\newcommand{\bra}[1]{\ensuremath{\left< #1\,\right|}}
\newcommand{\ket}[1]{\ensuremath{\left|\, #1\right>}}
\newcommand{\braket}[2]{\ensuremath{\left< #1\, |\, #2\right>}}
\newcommand{\commutator}[2]{\ensuremath{\left[#1,#2\right]}}
\newcommand{\cnot}{\ensuremath{\text{CNOT}}}
\newcommand{\biket}[2]{\ensuremath{\left|\, \begin{array}{*{#1}{c}} #2 \end{array} \right\rangle}}
\newcommand{\bibra}[2]{\ensuremath{\left\langle \, \begin{array}{*{#1}{c}} #2 \end{array} \right|}}
\newcommand{\tr}{\ensuremath{\mathrm{tr}\,}}		
\newcommand{\const}{\mathrm{constant}}

\journal{Annals of Physics}

\begin{document}

\begin{frontmatter}

\title{Entanglement propagation and typicality of measurements in the quantum Kac ring}

\author{Johannes M. Oberreuter, Ingo Homrighausen, Stefan Kehrein}

\address{Georg-August-Universit\"{a}t G\"{o}ttingen, Institute for Theoretical Physics, \\ Friedrich-Hund-Platz 1, 37077 G\"{o}ttingen, Germany}

\begin{abstract}
We study the time evolution of entanglement in a quantum version of the Kac ring. Our model consists of two spin chains and quantum gates instead of the classical markers. The gates take one qubit from each ring at a time as an input and entangle them. Subsequently, one ring is rotated. This protocol creates non-trivial entanglement between the two rings, which we measure by the entanglement entropy. The features of the entanglement evolution can best be understood by using knowledge about the behavior of an ensemble of classical Kac rings. For instance, the recurrence time of this quantum many-body system is twice the length of the chain and ``thermalization'' only occurs on time scales much smaller than the dimension of the Hilbert space. The model thus elucidates the relation between distribution of measurement results in quantum and classical systems: While in classical systems repeated measurements are performed over an ensemble of systems, the corresponding result is obtained by measuring the same quantum system prepared in an appropriate superposition repeatedly.
\end{abstract}

\end{frontmatter}

\section{Introduction}

How does the arrow of time arise? How does a system attain a thermal equilibrium state? And how can we apply the lessons learned from the thermalization of classical systems to quantum systems? 

Those are long-standing questions, whose origin lies in the fact that while the macroscopic dynamics, governing a system usually tends to some (thermal) equilibrium, the microscopic dynamics giving rise to it are usually time reversal invariant. Accordingly, it has been noted that every finite system must eventually return to its initial state once it has exhausted all its possibilities. It then shows \emph{Poincar\'{e} recurrence}. Usually, the time asymmetry arises when \emph{coarse graining} the microscopic theory. Thereby one usually makes the \emph{molecular chaos assumption}, which effectively amounts to neglecting existing correlations in the system. On time scales, which are not too long, the model behaves typical and thermalizes according to the Boltzmann equation. 

One accessible and pedagogical model to test the assumptions and the range of validity of this treatment is the Kac ring \cite{Kac1956} (We review this model in section \ref{sec:classkac}). It is a finite size model of classical particles whose dynamics is governed by a simple Hamiltonian, which switches one property of the particles back and forth.  Surprisingly, the recurrence time is here algebraic in the system size and not exponential. This puts strong bounds on the time scale, during which the Boltzmann equation is valid. 

Underlying classical evolution is a time-dependent quantum many body system. Any macroscopic equilibration process must be supported or at least allowed for by the microscopic laws of evolution. Therefore, understanding the time evolution of quantum many body systems is crucial for understanding macroscopic equilibration processes.

Recently, the question of thermalization in quantum systems, has gained considerable interest. The question of how quantum systems attain equilibrium is even more puzzling, since the evolution in an isolated quantum system is unitary
$$
\ket{\psi(t)} = \mathrm{e}^{-\mathrm{i} \widehat{H}t / \hbar} \ket{\psi(0)} = \sum_{\alpha} C_{\alpha} \mathrm{e}^{-\mathrm{i}E_{\alpha}t/\hbar} \ket{\alpha} \;, \quad \textrm{where} \quad \ket{\psi(0)} = \sum_{\alpha} C_{\alpha} \ket{\alpha}
$$
which means that the evolution of a state can be determined by the evolution of each energy eigenstate and a stationary thermal distribution can strictly speaking not be attained. 

In this paper, we propose a quantum many body system derived from the classical Kac ring to study questions around quantum thermalization in a manner as accessible and pedagogical as for the classical case. 
It consists of two parts, one of which can be regarded as the (laboratory) ``system'', the other as the ``environment''. The question to be examined is if and how the system reaches a (thermal) equilibrium state with respect to the environment, in other words, can parts of the system act as its heat bath?
Just as the in classical model, the evolution of the system is governed by a very simple protocol. As a typical quantum property it creates entanglement between two parts of the system, whereupon the same action can also destroy entanglement. Thus, as a new measurable quantity, this model has non-trivial entanglement entropy, which has no classical interpretation. Therefore, our model is truly quantum many body in nature. The creation and destruction of entanglement between parts of the system is a feature that previous quantum realizations of the Kac ring lack  \cite{Tavernier1976,Roeck2003}.
The natural expectation is that under non-trivial dynamics, the entanglement entropy will grow and finally saturate. We will explain that this is not quite the case here, since the reversal of entanglement creation and full recurrence also lead to a decrease of the entanglement entropy. 

This foremost pedagogical model allows to study explicitly the validity of (the analogue of) the quantum Boltzmann equation, which arises because of the oblivion of the correlations in the system. The short recurrence time makes the anti-Boltzmann behavior easily accessible, whilst the timescales necessary to do that are very hard to reach in numerical simulations of real world
systems.

Another question which is generally hard to answer in statistical mechanics is, how typical the solution of the Boltzmann equation is for a specific ensemble of systems or how typical the result of a measurement is in the quantum case. To quantify this, one needs to calculate the variance of the distribution, for which an exact solution is usually necessary. This is usually impossible for models which show Boltzmann-like behavior because these are non-integrable. This makes the quantum Kac ring such an interesting toy model.

It is instructive to compare the classical and quantum versions of the Kac ring with respect to their statistical properties. Our system is constructed such that the quantum system can be seen as a superposition of a large number of classical configurations. It is thus a parallel computing device, just like an ordinary quantum computer, which evaluates the classical evolution of several systems in parallel. For an ensemble of classical systems, the typical behavior can only be measured as an average over a large number of different systems in the ensemble. In the quantum version, however, this is automatically done and an individual system already behaves typical in general. The prediction here is the expectation value for a specific operator, which makes averaging obsolete. This means that the typical behavior of an ensemble of classical Kac rings will be reproduced by a number of measurements performed on only one quantum system. For this, a single pure state is sufficient; it is not necessary to use a mixed state. 

This paper is organized as follows. In section \ref{sec:classkac}, we review the classical Kac ring model, following closely the review \cite{Gottwald2009}. In section \ref{sec:qkac}, we present the extension of the model into the quantum regime. In section \ref{sec:observables} we present two interesting quantum observables in the quantum Kac ring, namely the entanglement entropy and the magnetization as well as their time development. We explain in section \ref{sec:typicality} that the evolution of the quantum model can be best understood by the relation between its measurements and statistical properties to the classical model. We close with concluding remarks in section \ref{sec:conclusions}.

\section{A brief review of the classical Kac ring model}
\label{sec:classkac}

\begin{figure}
\centering
\includegraphics[width=0.3\textwidth]{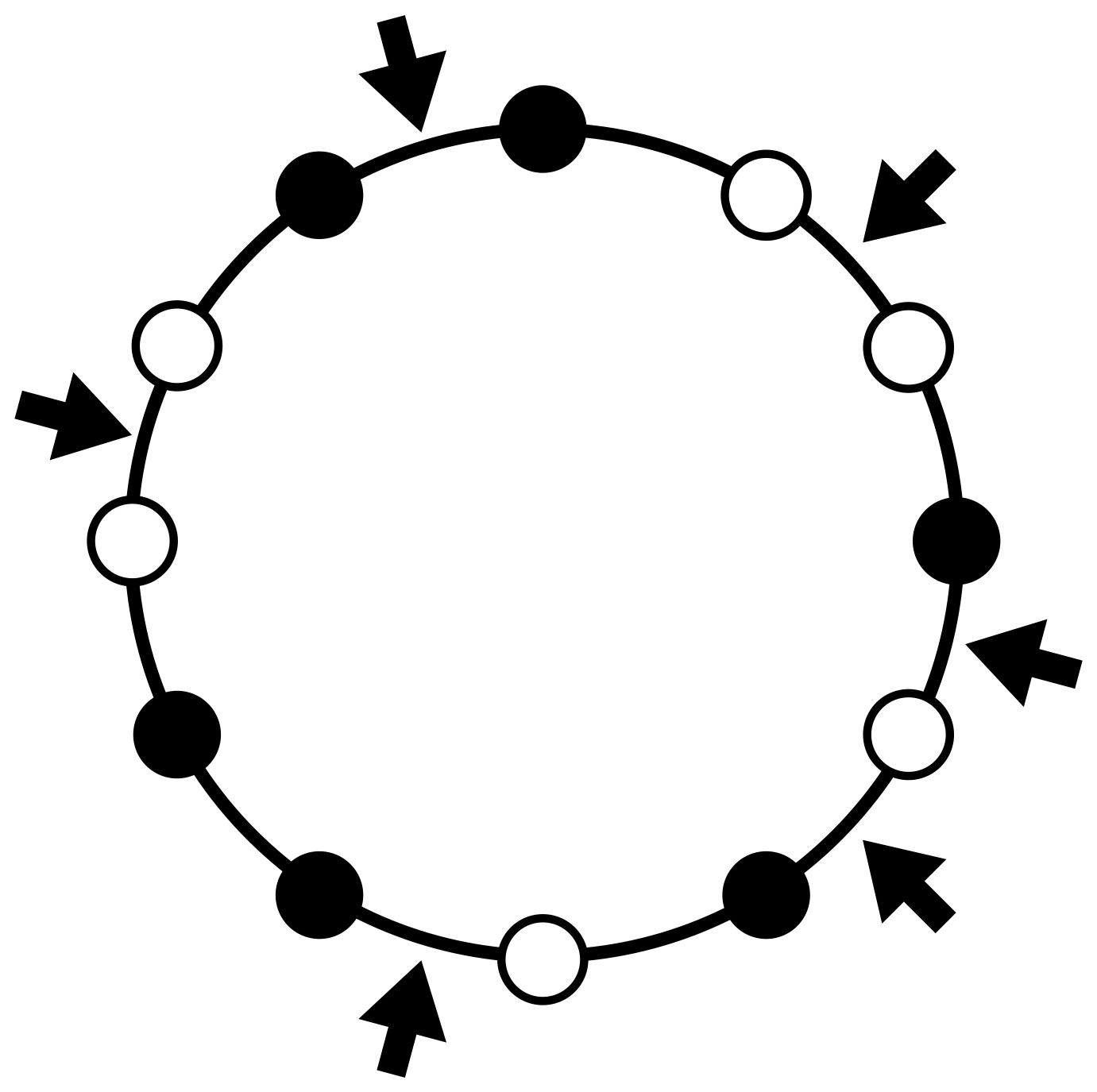}
\caption{An illustration of the Kac ring model. The balls can be in the states black or white. The markers are depicted by arrows on the bonds.}
\label{fig:clasKac}
\end{figure}

The Kac ring is a simple statistical many-body model, in which Boltzmann behavior as well as anti-Boltzmann behavior and recurrence can be studied \cite{Kac1956}. The degrees of freedom in the model are $N$ balls on a ring, which can each be in either of the states \emph{white} and \emph{black}. Furthermore, a number $n$ of markers is distributed over the different bonds between the individual balls (cf. Fig.~\ref{fig:clasKac})

The time development in the ring consists of two operations. At every discrete time step, the ring is rotated by one place, while the markers remain fixed. Furthermore, if a ball crosses a marker during one step of the evolution, it switches color, so a black ball becomes white and a white ball turns black.

Although the precise dynamics of the model is quite complicated, there is an immediate observation one can make: After an even number of flips, every ball is back to its initial state. Hence, no matter how the markers and initial colors are distributed, every system will recur to its initial state after $N$ time steps, if the total number of markers $n$ is even or after $2N$ time steps, if the number of markers is odd.

\subsection{Molecular chaos assumption}

The following treatment is borrowed from the review \cite{Gottwald2009}. One can derive the equilibration of the system according to some ``Boltzmann equation'' from a molecular chaos assumption. 

Let us denote the number of black and white balls at a given time step $t$ with $B(t)$ and $W(t)$, respectively. Besides, we denote the number of black balls, which are on a place just before crossing a marker, such that their conversion to white balls is imminent on the next step of time evolution by $b(t)$ and $w(t)$ for such white balls. Then, the number of black and white balls after the next time step can be written as
\begin{eqnarray*}
B(t+1) &= &B(t) + w(t) -b(t) \\
W(t+1) &= &W(t) + b(t) -w(t) \;.
\end{eqnarray*}
We now consider the ``grayness'' 
$\Delta(t) = B(t) - W (t)$ of the system as an observable. In a more physical sense, this observable could be interpreted as the magnetization of the system, namely the difference between spins up and spins down. At the next time-step, the value of this observable is easily determined to be $\Delta(t+1) = \Delta (t) + 2 w(t) - 2 b(t)$.

The molecular chaos assumption now enters as the approximation, that the ratio of black balls to turn white and the total number of black balls (just as the same ratio for the white balls), is more or less the same as the probability of finding a marker on a bond
$$
\mu = \frac{n}{N} = \frac{b}{B} = \frac{w}{W} \;.
$$
With this assumption, the time evolution of the grayness can now be written as
$$
\Delta(t+1) = \Delta(t) + 2 \mu W(t) - 2\mu B(t) = (1-2\mu) \Delta(t) 
$$
and the recursion can be continued all the way to the initial configuration. The Boltzmann equation then reads
\begin{equation}
\label{eq:Boltzmann}
\Delta(t) = (1-2\mu)^t \Delta(0) \;.
\end{equation}
We see that for $\mu < \frac12$, the value of $\Delta$ strives to its equilibrium value naught.

This, however, is in contradiction to the earlier observation, that the initial configuration of the system will recur after at least $2N$ time steps. This signals, that the molecular chaos assumption breaks down.

\subsection{Microscopic dynamics}
\label{sec:micdyn}

We follow again the presentation given in \cite{Gottwald2009}. To determine the grayness $\Delta(t)$ from the microscopic configuration of the system, we need to determine the probability that the color of a specific ball is switched at every time. Let $\chi_i(t)$ be the color of the $i$\textsuperscript{th} ball at time $t$, where $\chi_i$ takes the values $1$ for black and $-1$ for white. The color is flipped if there is a marker on the $i$\textsuperscript{th} site, which is denoted by $m_i=-1$. The color remains the same if the value of $m_i=1$. Then, $\chi_{i+1}(t+1) = m_i \chi_i(t)$. After $t$ time steps, a ball has passed $t$ consecutive markers and we can write
$$
\Delta(t) = \sum_{i=1}^N \chi_i = \sum_{i=1}^N m_{i-1} m_{i-2} \dots m_{i-t} \chi_{i-t}(0) \;.
$$

To calculate the evolution of the average we can now simply average over the markers' positions if the initial configuration of colors is supposed not to vary between different systems. We obtain
$$
\langle \Delta (t) \rangle = \sum_{i=1}^N \langle m_{i-1} m_{i-2} \dots m_{i-t} \rangle \chi_{i-t}(0) \;.
$$
For the average of the marker values, the starting point does not make a difference and we can relabel the markers to obtain
\begin{equation}\label{eq:microderivBoltzmann}
\langle \Delta (t) \rangle = \langle m_1 m_2 \dots m_{t} \rangle \sum_{i=1}^N \chi_{i-t}(0) = \langle m_1 m_2 \dots m_{t} \rangle  \Delta (0) \;.
\end{equation}
We now observe, that for the first run $0 \leq t \leq N$, there are no periodicities and all the markers are independent of each other. Their product is $-1$ and $1$ for an odd and even number of markers, respectively. The average over the product can thus be written as
$$
\label{eq:markeraverage}
 \langle m_1 m_2 \dots m_{t} \rangle = \sum_{j=0}^t (-1)^j p_j (t) \;,
$$
where $p_j(t)$ is the probability of finding exactly $j$ markers on $t$ consecutive sites. The latter is given by a binomial distribution 
$$
p_j (t) = \binom{t}{j} \mu^j (1-\mu)^{t-j} 
$$
due to the independence of the markers. These probabilities can now be summed up to give
$$
\langle m_1 m_2 \dots m_{t} \rangle = \sum_{j=0}^t  \binom{t}{j} (-\mu)^j (1-\mu)^{t-j} = (1-2\mu)^t
$$
and we thus re-obtain the Boltzmann equation \eqref{eq:Boltzmann} 
 
Note, however, that the assumption of independence does not hold any more for the second run, $N \leq t \leq 2N$, in which case the balls will pass at least some markers twice. We can account for these periodicities by first observing
$$
\langle m_1 m_2 \dots m_{t} \rangle = \langle m_{t+1} m_{t+2} \dots m_{2N} \rangle =\langle m_1 m_2 \dots m_{2N-t} \rangle 
$$
which is due to the $N$-periodicity of the lattice and subsequent index shift. If we use the last relation in \eqref{eq:microderivBoltzmann}, we find
$$
\langle \Delta (t) \rangle_{t > N} =  (1-2\mu)^{2N-t} \Delta (0) \;.
$$
This equation describes the \emph{anti-Boltzmann} behavior, since the exponent decreases with time. For $t \to 2N$, the observable $\Delta$ has returned to its initial value.

\section{The quantum Kac ring}
\label{sec:qkac}

\begin{figure}
\centering
\includegraphics[width=0.4\textwidth]{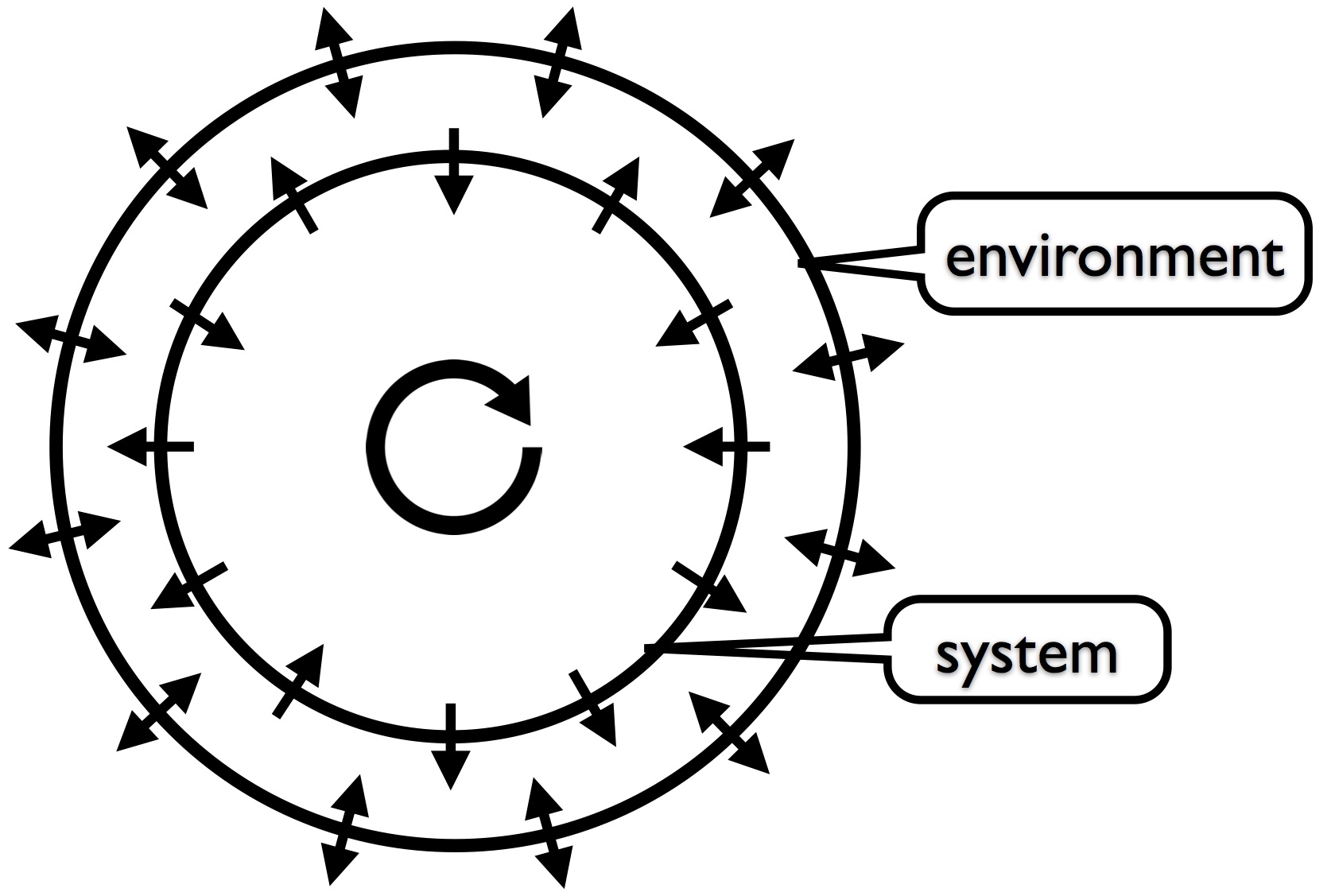}
\caption{An illustration of our realization of the quantum Kac ring. It consists of two adjacent rings, one of which is rotated at each time step. We refer to one ring as the \emph{system}, which is controlled directly and to the other as \emph{environment}, with which the system gets entangled by direct interactions.}
\label{fig:quantKac}
\end{figure}

We generalize this classical model to a quantum version, in which entanglement can be generated and can spread through the system as well as be destructed by one and the same time evolution. Quantum realizations of the Kac ring, in which spins play the role of the degrees of freedom and a spin-flip corresponds to the change of colors have been treated earlier \cite{Tavernier1976,Roeck2003}. However, those models did not feature dynamical creation and destruction of entanglement.
To obtain this, we are considering two adjacent spin chains with each $N$ sites. Hence, the Hilbert space is $\mathcal{H} = \bigotimes_{i=1}^N ( \mathbb{C}^2 \otimes \mathbb{C}^2)$. In our model, the markers are now quantum operations on \emph{each} site, where they govern the interaction and creation of entanglement between the two adjacent spins on the same site in the two chains (cf. Fig.~\ref{fig:quantKac}).

\begin{figure}
\centering
\def\svgwidth{0.4\textwidth}
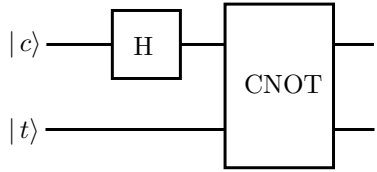
\caption{A CNOT operation acting on two qubits, where the control qubit has been rotated into a superposition $\ket{0}\pm \ket{1}$ generates all the four Bell-states.}
\label{fig:cnotentangling}
\end{figure}

A standard protocol for entangling two qubits is to perform a \emph{controlled-not} (CNOT) operation onto them, after one of them has been prepared into a superposition (see. Fig.~\ref{fig:cnotentangling})
The CNOT gate flips the ``target'' bit if and only if the ``control'' bit is true, whereby the control bit remains unchanged. In other words
\begin{equation}
\cnot(\ket{c} \otimes \ket{t}) = \ket{c} \otimes \ket{(t + c) \bmod{2} } \;.
\end{equation}
As an operator, it can be represented in matrix notation as
\begin{equation}
\cnot = \begin{pmatrix} 1&0 & &\\0&1 & &\\ & & 0&1\\ & & 1&0 \end{pmatrix} \;.
\end{equation}
Since the quantum gate does not change the control bits it appears appropriate to regard them as the environment whereas the system consists of the target bits.

The superposition of the control-bit can be achieved by performing a Hadamard gate, which is defined by
\begin{equation}
H = \frac1{\sqrt{2}} \begin{pmatrix} 1&1\\1& -1 \end{pmatrix} \;.
\end{equation}
By combining the four different possible combinations of values for the two qubits, one can herewith generate all the four Bell-states, which form a basis of the two-bit Hilbert space. The corresponding operations are
\begin{eqnarray}
\ket{\Phi^+} &=& \cnot ( H \ket{0} \otimes \ket{0}) = \frac1{\sqrt{2}} ( \ket{00} + \ket{11} ) \;, \\
\ket{\Phi^-} &=& \cnot ( H \ket{1} \otimes \ket{0}) = \frac1{\sqrt{2}} ( \ket{00} - \ket{11} ) \;, \\
\ket{\Psi^+} &=& \cnot ( H \ket{0} \otimes \ket{1}) = \frac1{\sqrt{2}} ( \ket{01} + \ket{10} ) \;, \\
\ket{\Psi^-} &=& \cnot ( H \ket{1} \otimes \ket{1}) = \frac1{\sqrt{2}} ( \ket{01} -  \ket{10} ) \;.
\end{eqnarray}
Since these four states are a basis, all statements about the quantum Kac ring can be deduced in terms of them.

The behavior of the quantum Kac ring becomes interesting due to the relative rotation between the rings. The quantum gate imitatates the property of the classical markers that they reverse their own action: a marker makes a passing black ball switch color to white and back to black upon the next action. 
The same property holds for the $\cnot$ gate since it squares to the identity
\begin{equation}
\cnot^2 = \mathbbm{1} \;.
\end{equation}

Because of the behavior of the individual markers, the recurrence time in the classical Kac ring model is $N$ or $2N$ for an even or odd number of markers, respectively. A similar property is reproduced by the quantum Kac ring, if the operations acting on the individual sites commute
\begin{equation}
\commutator{\cnot_i^j}{\cnot_i^l} = 0 \;,
\end{equation}
where the subscript denotes the site in the control ring which the gate takes as input together with the j\textsuperscript{th} and l\textsuperscript{th} target bit, respectively. 
That this is the case becomes immediately apparent if one decomposes the system into the $\{ \ket{0}, \ket{1} \}$ basis. 
Each basis vector represents a classical Kac ring, the markers of which are situated on the sites where the control bits are $\ket{1}$. The spin flip corresponds to the changing of color in the classical analogue. Then, a spin which has passed an even number of control bits of value $\ket{1}$ has reverted to its original color. Hence, for every basis vector, the recurrence time is $N$ and $2N$ for an even and odd number of $\ket{1}$s in the control ring, respectively. Since this holds for every basis vector, all entanglement in the full system is destroyed at least after $2N$ steps.

\subsection{Many-body notation}

One simplifying feature of our model is the fact that the control ring never changes. It neither gets rotated nor do any of its spins get flipped. Furthermore, every site of the control ring interacts with only one site of the target ring at every given time. Therefore, it is convenient to introduce the notation
\begin{equation}
\bigotimes_{i=1}^N \ket{c_i} \otimes \ket{t_i} = \biket{4}{c_1 & c_2 & \dots & c_N\\ t_1 & t_2 & \dots & t_N} \;.
\end{equation}
Every target bit is denoted below the control bit, which determines if it gets flipped or not. Of course, since the target ring is rotating at every step, the position of the target bits changes with time. The Rotation is implemented by
\begin{equation}
R \biket{4}{c_1 & c_2 & \dots & c_N\\ t_1 & t_2 & \dots & t_N} = \biket{4}{c_1 & c_2 & \dots & c_N\\ t_N & t_1 & \dots & t_2} \;.
\end{equation}
For every qubit that is in a superposition of two states, the number of kets needed to describe that state is doubled. Hence, if all $N$ control bits are in a superposition, the state of the system is described by $2^N$ kets with distinct distributions of $0$ and $1$ qubits over the control ring sites. This number would increase even more, if we were to allow superpositions for the target bits, too, but this would complicate matters unnecessarily, since to generate entanglement, it is sufficient to have the control qubit in a superposition. Note, however, that every constituent of the superposition of the control ring is by itself a marker distribution of a classical Kac ring.

In practice, one wants not only to have an even superposition of $\ket{0}_c$ and $\ket{1}_c$, where the coefficients are equal, but to allow also for relative coefficients corresponding to a general point on the Bloch sphere. If the coefficients quantifying the individual contributions to a superposition are not trivial, we denote them in front of the ket as is customary
\begin{equation}
\frac1{\sqrt{|a|^2 + |b|^2}} \left( a \ket{0}_c + b \ket{1}_c \right) \otimes \ket{t} = \frac1{\sqrt{|a|^2 + |b|^2}} \left(a \biket{1}{0\\t} + b \biket{1}{1\\t} \right)\;.
\end{equation}
Note that the complex coefficients are fixed at the position of their control bit. Every control bit can thus be written as
\begin{equation}
a \ket{0}_c + b \ket{1}_c = \left( a \atop 0 \right) + \left( 0 \atop b \right) \;.
\end{equation}
Since the overall normalization takes into account all constituents of the superposition, it is convenient to introduce
\begin{equation}
A = \sqrt{\prod_{i=1}^N \left(a_i^2 + b_i^2 \right)} \;.
\end{equation}
As a last remark, let us note that after a few steps of the evolution, the target qubit associated with the $\ket{0}_c$ and with the $\ket{1}_c$ control bit does not need to and in general will not be the same, since the CNOT gate acts differently on the target for different control bits. The states on one site of the combined rings will then look like 
\begin{equation}
a \biket{1}{0\\t} + b \biket{1}{1\\ \tilde{t}} \;.
\end{equation}

\subsection{Evolution in the Quantum Ring}
\label{sec:results}

Let us look at a simple setup with $N=2$ for a start. Here, we bring the control ring into a specific state by applying the Hadamard-gate to $\ket{00}_c$ and obtain $H \ket{00}_c  = \frac1{\sqrt{2}} (\ket{0}+\ket{1})_{c_1} \otimes  \frac1{\sqrt{2}} (\ket{0}+\ket{1})_{c_2}$ and a target ``ring'', where the qubits are in the state $\ket{01}_{t}$. In our notation, the initial state, in which we prepare the system looks like
\begin{equation}
\ket{\psi(0)} = \frac12 \left(
\biket{2}{0&0\\0&1} + \biket{2}{0&1\\0&1} + \biket{2}{1&0\\0&1} + \biket{2}{1&1\\0&1}\right) \;.
\end{equation}
to which we apply the CNOT-operation between both pairs of sites as described above. The resulting state
\begin{equation}
\label{eq:psi1}
\ket{\psi(1)} = \frac12 \left( \biket{2}{0&0\\0&1} + \biket{2}{0&1\\0&0} + \biket{2}{1&0\\1&1} + \biket{2}{1&1\\1&0}\right) \;,
\end{equation}
 is entangled between the two rings. Note that the first line of the ket, which describes the state of the control ring, has not changed for any term in \eqref{eq:psi1}. The entanglement between the two rings can be seen by inspection for that simple model. Now we rotate the target ring by one site and subsequently apply the CNOT-gates again. The state of the whole system at the second step is
 \begin{eqnarray*}
 \label{eq:2siteringevolution}
 \ket{\psi(2)} &=& ( \mbox{CNOT}_1 \otimes \mbox{CNOT}_2 ) R \ket{\psi(1)} \\
 &=& \frac12 ( \mbox{CNOT}_1 \otimes \mbox{CNOT}_2 ) \\ & &  \left( \biket{2}{0&0\\1&0} + \biket{2}{0&1\\0&0} + \biket{2}{1&0\\1&1} + \biket{2}{1&1\\0&1}\right) \\
 &=& \frac12 \left( \biket{2}{0&0\\1&0} + \biket{2}{0&1\\0&1} + \biket{2}{1&0\\0&1} + \biket{2}{1&1\\1&0}\right)
 \end{eqnarray*}
 The same protocol is now repeated over and over again and we find
\begin{eqnarray}
\ket{\psi(3)} &=& \frac12 \left( \biket{2}{0&0\\0&1} + \biket{2}{0&1\\1&1} + \biket{2}{1&0\\0&0} + \biket{2}{1&1\\1&0}\right) \;, \\
\ket{\psi(4)} &=& \frac12 \left( \biket{2}{0&0\\1&0} + \biket{2}{0&1\\1&0} + \biket{2}{1&0\\1&0} + \biket{2}{1&1\\1&0}\right)\\&=& R^{-1} \ket{\psi(0)} \;.
\end{eqnarray}
We see that upon rotation in the target ring, the last step shows recurrence $R \ket{\psi(4)} = \ket{\psi(0)}$ after $2N=4$ steps as expected.

\subsection{Eigenvectors and Eigenvalues of the Time Evolution}

The standard approach to examine the behavior of a quantum system is to diagonalize its Hamiltonian and determine the eigenstates of the system. It is, however, also a general observation that for the dynamics of a quantum many body system, knowledge of the eigenvectors is of only limited use. 

We shall demonstrate this briefly. The time evolution is composed of the CNOT operation and subsequent rotation. As an example, we consider the eigenvectors of the evolution for the $N=2$ system, for which it can be represented as
\begin{equation}
R \cdot \cnot \otimes \cnot \;,
\end{equation}
where the rotation $R$ can be decomposed into appropriate permutations.

We find eight eigenvectors with eigenvalue 1, namely
\begin{eqnarray}
& \biket{2}{0&0\\0&0}\;, \quad \biket{2}{0&0\\0&1} \;, \quad  \biket{2}{1&1\\1&1}+\biket{2}{1&1\\0&0} \;, \\ 
& \biket{2}{1&0\\0&0} + \biket{2}{1&0\\0&1} + \biket{2}{1&0\\1&0} + \biket{2}{1&0\\1&1} \;, \quad  \biket{2}{1&1\\0&1} \\
& \biket{2}{0&1\\0&0} + \biket{2}{0&1\\0&1} + \biket{2}{0&1\\1&0} + \biket{2}{0&1\\1&1} \;, \quad  \biket{2}{0&0\\1&1}  \;, \quad \biket{2}{0&0\\1&0} \;.
\end{eqnarray}
There are also four eigenvectors for the eigenvalue $-1$, namely
\begin{eqnarray}
\biket{2}{1&1\\1&1}-\biket{2}{1&0\\1&1} \;, \quad \biket{2}{1&0\\0&0} - \biket{2}{1&0\\0&1} + \biket{2}{1&0\\1&1} - \biket{2}{1&0\\1&0} \;, \\
\biket{2}{0&1\\1&1} - \biket{2}{0&1\\1&0} + \biket{2}{0&1\\0&0} - \biket{2}{0&1\\0&1} \;, \quad \biket{2}{0&0\\1&0} - \biket{2}{0&0\\0&1} \;.
\end{eqnarray}
Finally, there are two eigenvectors which entail a phase shift with an eigenvalue $\im$
\begin{eqnarray}
\biket{2}{1&0\\1&1} - \im \biket{2}{1&0\\1&1} - \biket{2}{1&0\\0&0} + \im \biket{2}{1&0\\0&1} \;, \\ \quad \biket{2}{0&1\\1&1} + \im \biket{2}{0&1\\1&0} - \biket{2}{0&1\\0&0} - \im \biket{2}{0&1\\0&1} \;,
\end{eqnarray}
and two with an eigenvalue $-\im$
\begin{eqnarray}
\biket{2}{1&0\\1&1} + \im \biket{2}{1&0\\1&0} - \biket{2}{1&0\\0&0} - \im \biket{2}{1&0\\0&1} \;, \\ \quad \biket{2}{0&1\\1&1} - \im \biket{2}{0&1\\1&0} + \biket{2}{0&1\\0&0} - \im \biket{2}{0&1\\0&1} \;.
\end{eqnarray}

The recurrence after four time steps is reflected in the fact that the eigenvalues are fourth roots of unity.
However, other essential properties like the generation of entanglement are not obvious from the eigenstates and --values of the system.

\section{Observables}
\label{sec:observables}

\subsection{Entanglement entropy}

An often useful measure for the amount of entanglement is the entanglement entropy. For a bi-bipartite system consisting of two parts $A$ and $B$, the entanglement entropy of one subsystem $A$, say, is defined as
\begin{equation}
S_A = - \tr \rho_A \log \rho_A = S_B
\end{equation}
where $\rho_A$ is the reduced density matrix of subsystem $A$ achieved by partially tracing over $B$
\begin{equation}
\rho_A = \mathrm{tr}_B \, \rho \;.
\end{equation} 
Note that the entanglement entropy is symmetric between the two subsystems $A,B$.

Let us focus on the case, where we are interested in the entanglement between the control ring and the target ring. In our notation, the density matrix of the full system is 
\begin{equation}
\rho=\frac1{A^2} \sum_{i,j} \biket{4}{c_1^i & c_2^i & \dots & c_N^i\\ t_1^i & t_2^i & \dots & t_N^i} \bibra{4}{c_1^j & c_2^j & \dots & c_N^j\\ t_1^j & t_2^j & \dots & t_N^j} \;,
\end{equation}
where we want to trace out the control ring, which we interpret as the environment. The reduced density matrix for the target ring is obtained by tracing over the first line
\begin{equation}
\rho_{\mathrm{target}} = \frac1{A^2} \sum_{i,j} \ket{ t_1^i  t_2^i  \dots  t_N^i} \bra{t_1^j  t_2^j  \dots  t_N^j} \braket{c_1^i c_2^i \dots c_N^i}{c_1^j c_2^j \dots c_N^j} \;.
\end{equation}
In the above sum, every vector $\ket{c_1^i \dots c_N^i}$ denotes a different configuration of qubits in the control ring, i.e. it corresponds to a  different basis vector. Since different (classical) configurations of the control ring are orthogonal to each other, only the $2^N$ diagonal terms of the density matrix can be non-vanishing.

Since the control ring never changes and neither do the coefficients associated to every constituent of its superposition, the individual coefficients remain fixed under the time evolution and only $N$ values for the evolution of every target string need to be recorded at every time step.

For a particular constituent of the superposition, which is characterized by its (time-invariant) configuration of the control ring, only the configuration of the target qubits changes with time. For instance, a state
\begin{equation}
\label{eq:targetstatewithcoefficients}
\frac{a_1 b_2  b_3 a_4 b_5}{A} \biket{5}{ 1 & 0 & 0 & 1 &0 \\ t_1 & t_2 & t_3 & t_4 & t_5}
\end{equation}
contributes
\begin{equation}
\frac{a_1a_1^{\ast} \cdot b_2 b_2^{\ast} \cdot b_3 b_3^{\ast} \cdot a_4 a_4^{\ast} \cdot b_5 b_5^{\ast}}{A^2}  \ket{t_1 t_2 t_3 t_4 t_5} \bra{t_1 t_2 t_3 t_4 t_5}
\label{eq:supoexample}
\end{equation} 
to the density matrix. In general, such a constituent of the superposition can be denoted by $I=\{i \in \{1,\dots,N\} | c_i = \ket{1} \}$ and its contribution to the density matrix looks like
\begin{equation}
\rho_I=\frac{1}{A^2} \prod_{i \in I} a_{i} \prod_{j\in \{1,\dots,N\} \setminus I}  b_{j} \ket{t_1  t_2  \dots  t_N}\bra{t_1  t_2  \dots  t_N}\;.
\end{equation}
For the example in \eqref{eq:supoexample}, we would have $I = \{ 1,4 \} $. There are $2^N$ such contributions since the power set of $\{1 \dots N \}$ has that many elements. 

\begin{figure}
\centering
\includegraphics[width=0.4\textwidth]{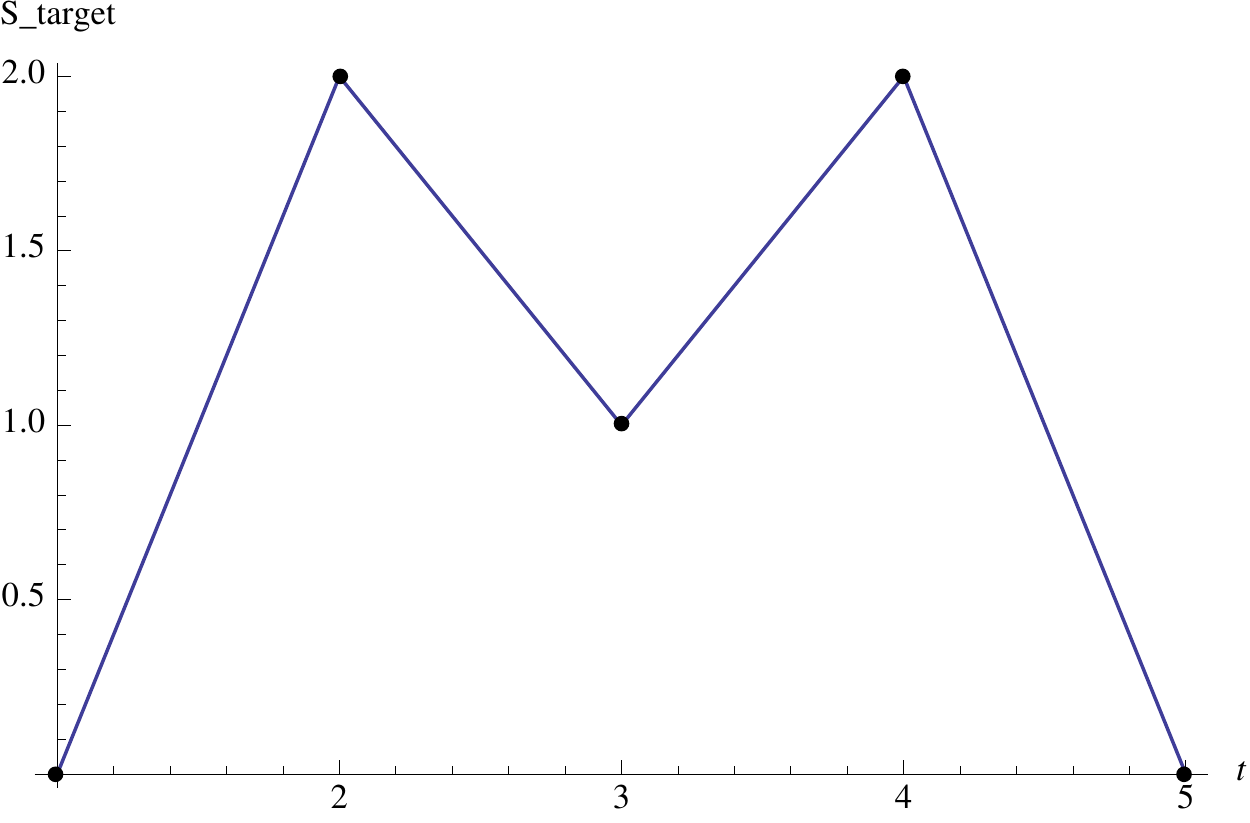}
\caption{The time evolution of the entanglement entropy of the target ring with the control ring traced out for an $N=2$ quantum Kac ring. The initial state of the target ring is $\ket{01}$. The same result holds for all the other possible initial configurations of the target ring. (The line is just a guide to the eye.)}
\label{fig:Nis2_Tis01}
\end{figure}

Let us examine the entanglement entropy for some specific systems. For the case of the $N=2$ system, whose time evolution we have been calculating in section \ref{sec:results}, the entanglement entropy can be computed easily. For the different time steps, we find the entanglement entropy $S_{\mathrm{target}}(t)$
\begin{equation}
S(1)= 0 \;, \quad S(2)=2 \;, \quad S(3)= 1 \;, \quad S(4)= 2 \;, \quad S(5) = 0 \;,
\end{equation}
where we have calculated the entropy with respect to basis $2$. This is a useful and customary choice for qubits. The evolution is depicted in Fig.~\ref{fig:Nis2_Tis01}.
The entanglement entropy is independent of the the initial configuration of the target ring. A different initial configuration will just lead to a different Bell state being created after the first CNOT operation. All Bell states have the same (maximal) entanglement entropy, which also holds for the subsequent time evolution. 

In the following, we are going to use the simple configuration $\ket{00 \dots 0}$ as initial configuration of the target ring. At first, we always use the maximal superposition for the control ring, which means that every control qubit is in a superposition $\ket{0}+\ket{1}$. In figure  \ref{fig:Nis5,10,11,15_entropy}, we depict the time evolution of the entanglement entropy with the control ring traced out for systems with $N=5, N=10, N=11$ and $N=15$.
\begin{figure}
\centering
\includegraphics[width=0.4\textwidth]{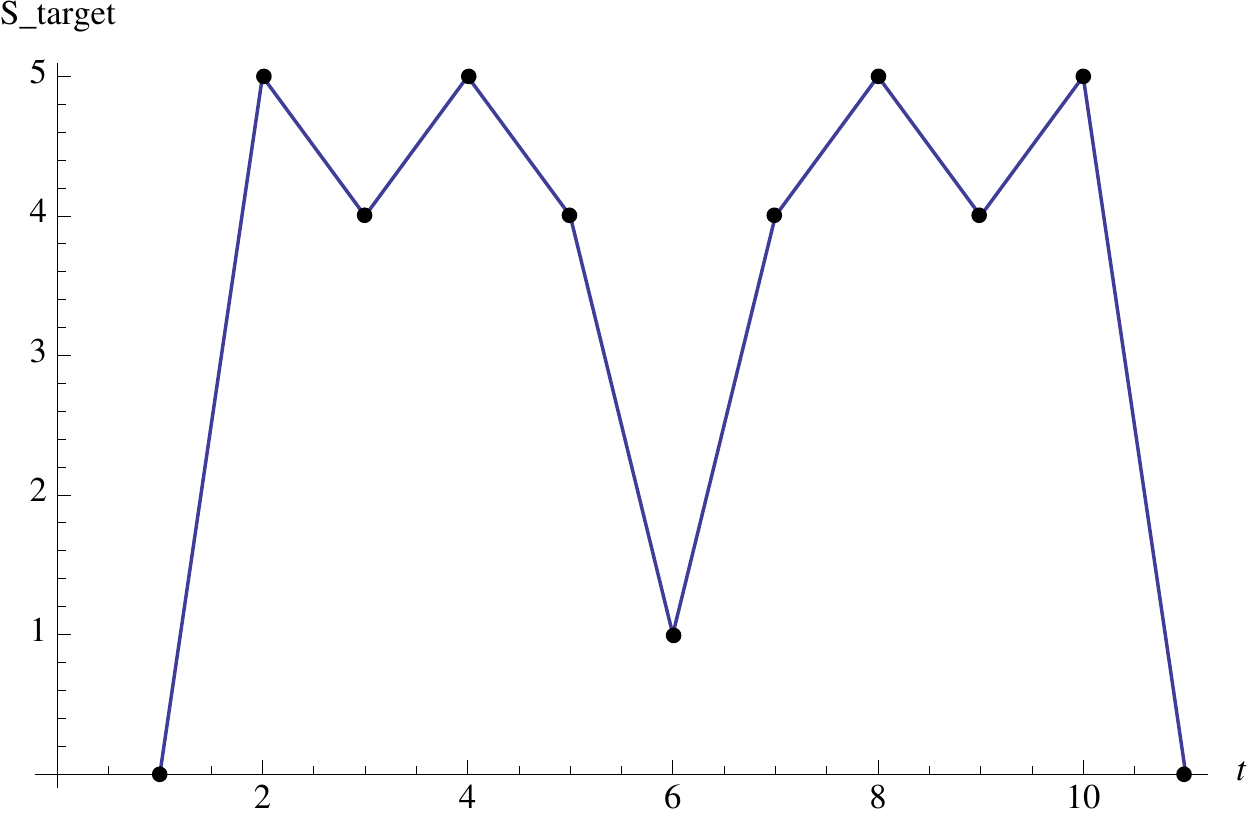}\hspace{0.1\textwidth}
\includegraphics[width=0.4\textwidth]{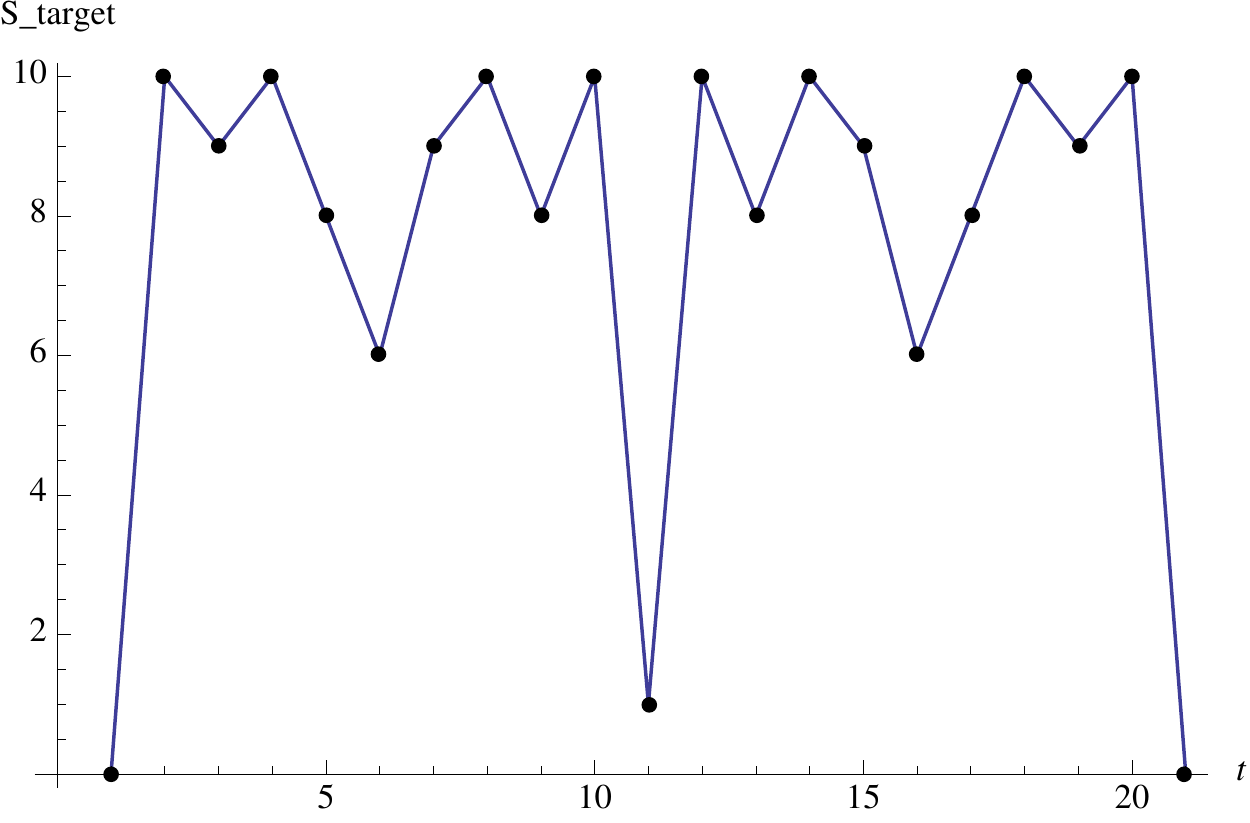}\\
\includegraphics[width=0.4\textwidth]{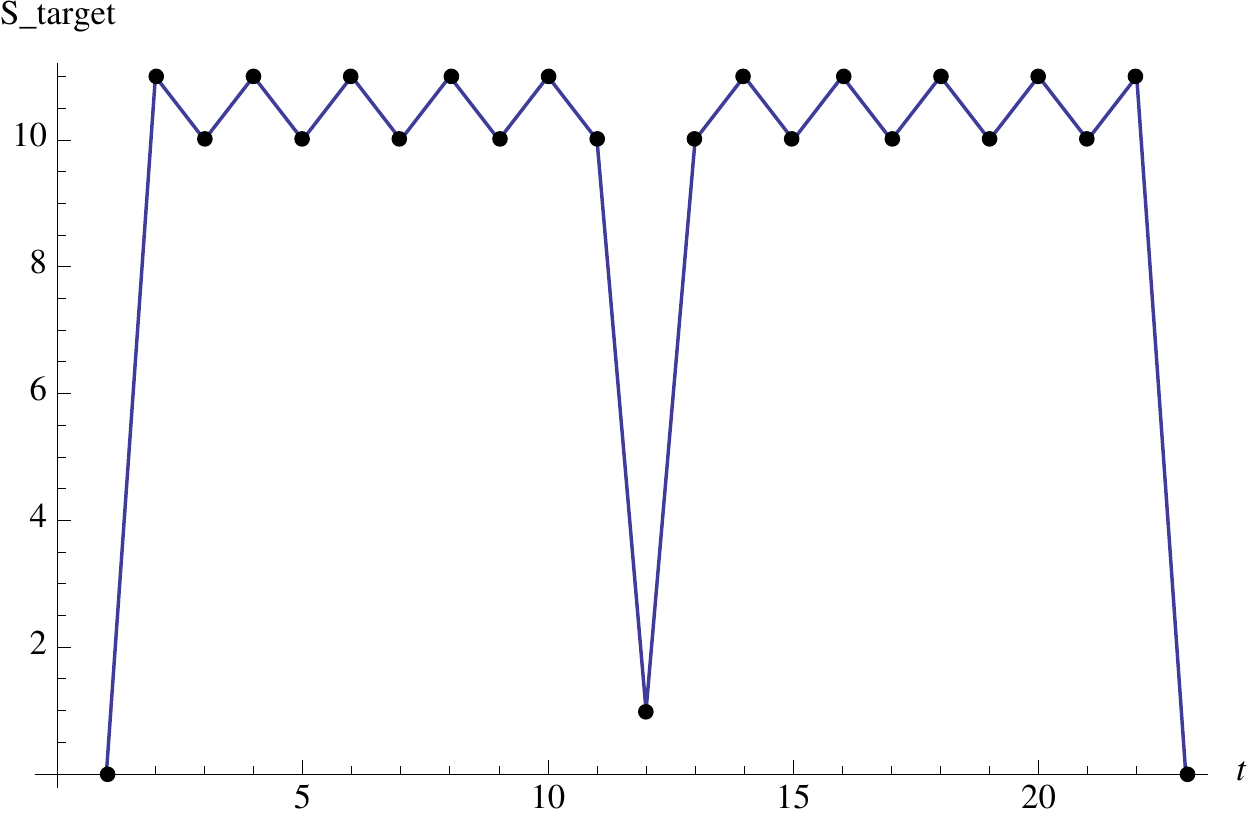}\hspace{0.1\textwidth}
\includegraphics[width=0.4\textwidth]{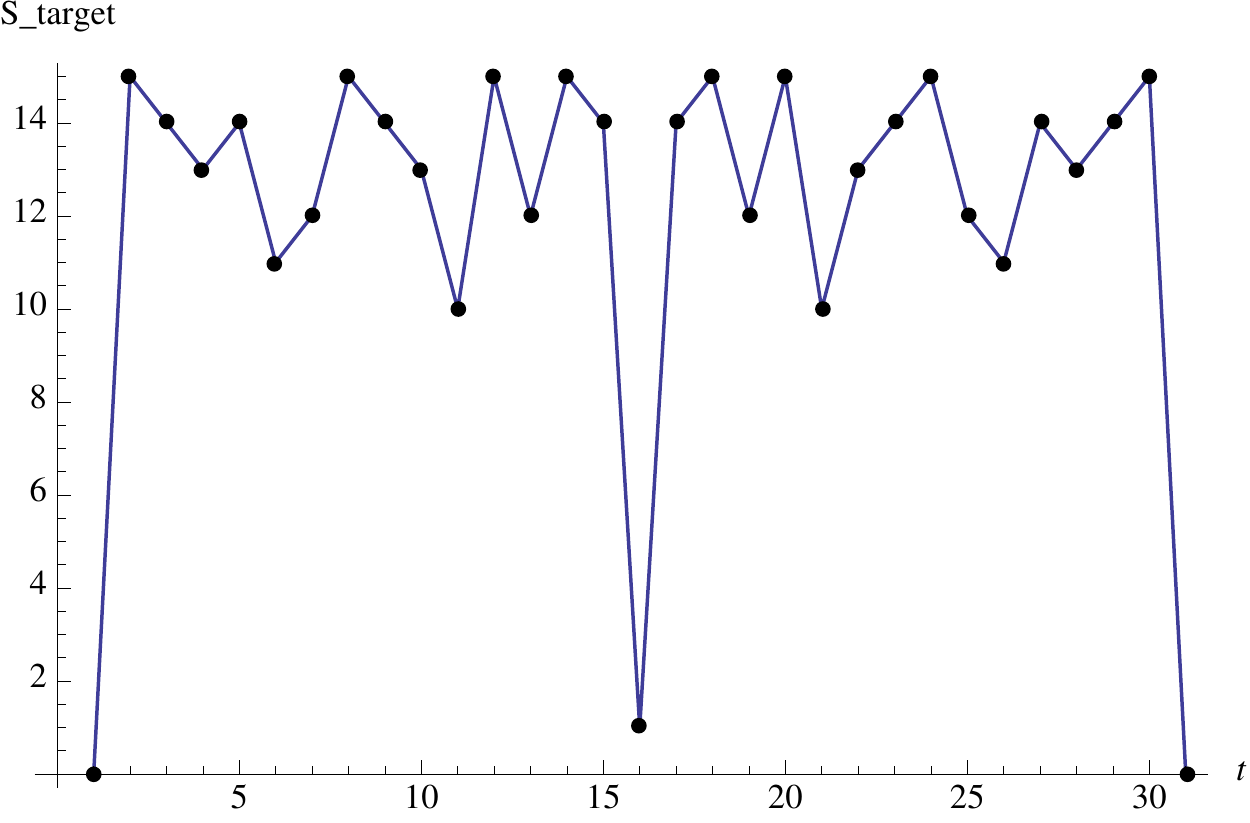}
\caption{The time evolution of the entanglement entropy of the target ring with the control ring traced out for an $N=5$ (top left), $N=10$ (top right), $N=11$ (bottom left) and $N=15$ (bottom right) quantum Kac ring. The initial state of the target ring is $\ket{00 \dots 0}$ and every qubit in the control ring is in a superposition for all cases. (The lines are just guides to the eye.)}
\label{fig:Nis5,10,11,15_entropy}
\end{figure}
We can make some interesting observations in these figures. An obvious remark is the recurrence of separability of the two rings after $t_{\mathrm{rec}}= 2N$ time steps, which comes from the recurrence of the exact configuration of the system. Another fact, which can be checked, is time reversal symmetry of the entanglement entropy, which reflects the time reversal symmetry of the whole evolution. 

An unexpected and curious feature is that in all the examples the entanglement entropy jumps to its maximum entropy immediately. Ordinarily one expects that the entanglement entropy grows with time and saturates eventually. This difference comes about because in the Kac ring, the two subsystems interact in the bulk, i.e. at every time step, every site in the system ring interacts with one site in the environment. In that way, every site in one subsystem can get entangled with a site in the other after one step only. In the usual setup, the subsystems only interact via the boundary and it therefore takes some time comparable to the size of the subsystem, before the maximum entanglement is created.

It is less obvious, why the minimum of the entanglement entropy appears to be attained after $N$ time-steps in all the models. The key to explaining this observation will give us an important insight into the relation between the quantum and the classical Kac ring models. The entanglement entropy is roughly proportional to the number of non-vanishing elements of the diagonal in the reduced density matrix. Every spot on the diagonal corresponds to a classical configuration of the target ring, of which there are $2^N$ different ones. We have seen that in our basis, only diagonal entries of the reduced density matrix can be non-vanishing. If we assume that all possible classical marker configurations are realized as a superposition in the control ring, as we did for generating the above plots, the number of classically available configurations bounds the entanglement entropy. Now recall, that after $N$ time steps, there are only two different states of the system possible classically: For an even number of markers, the system is completely reverted to its original state, whereas every spin has been flipped for an odd number of markers. This means that after $N$ steps, only two classical configurations are available and thus only two entries on the diagonal are non-vanishing. Thus, the entanglement entropy at this point is always unity but non-vanishing, which is the minimal non-trivial value.

\begin{figure}
\centering
\includegraphics[width=0.3\textwidth]{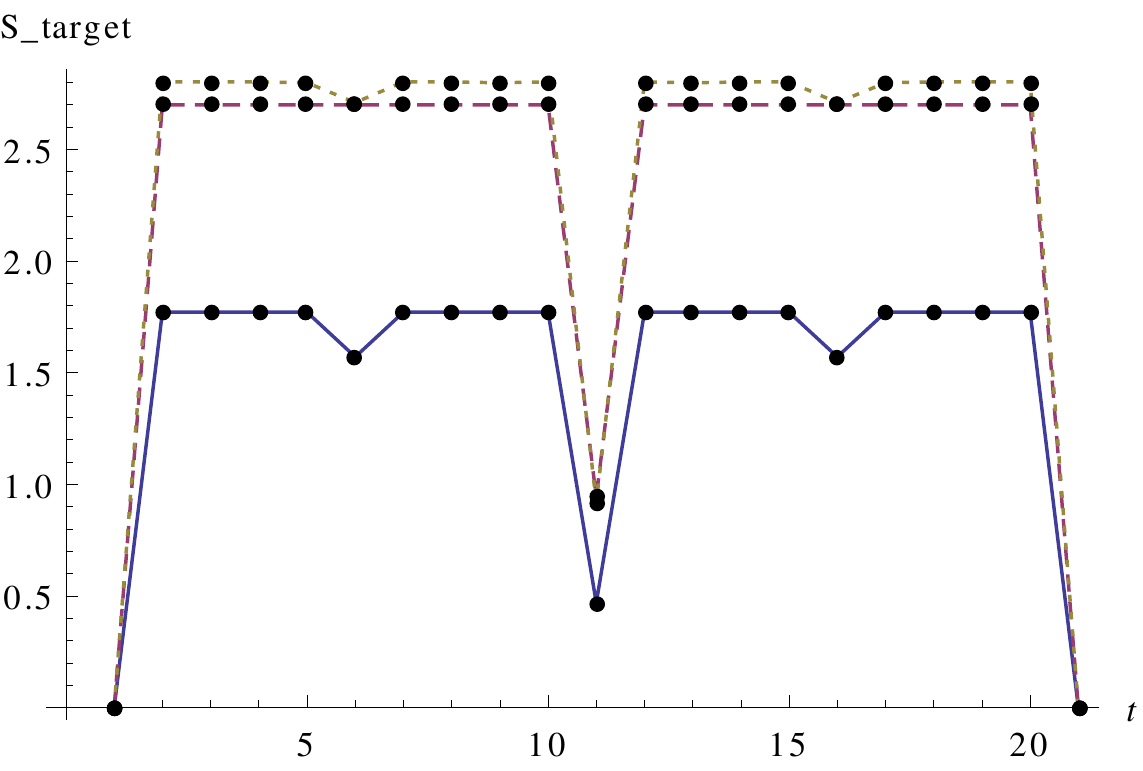} \hfill \includegraphics[width=0.3\textwidth]{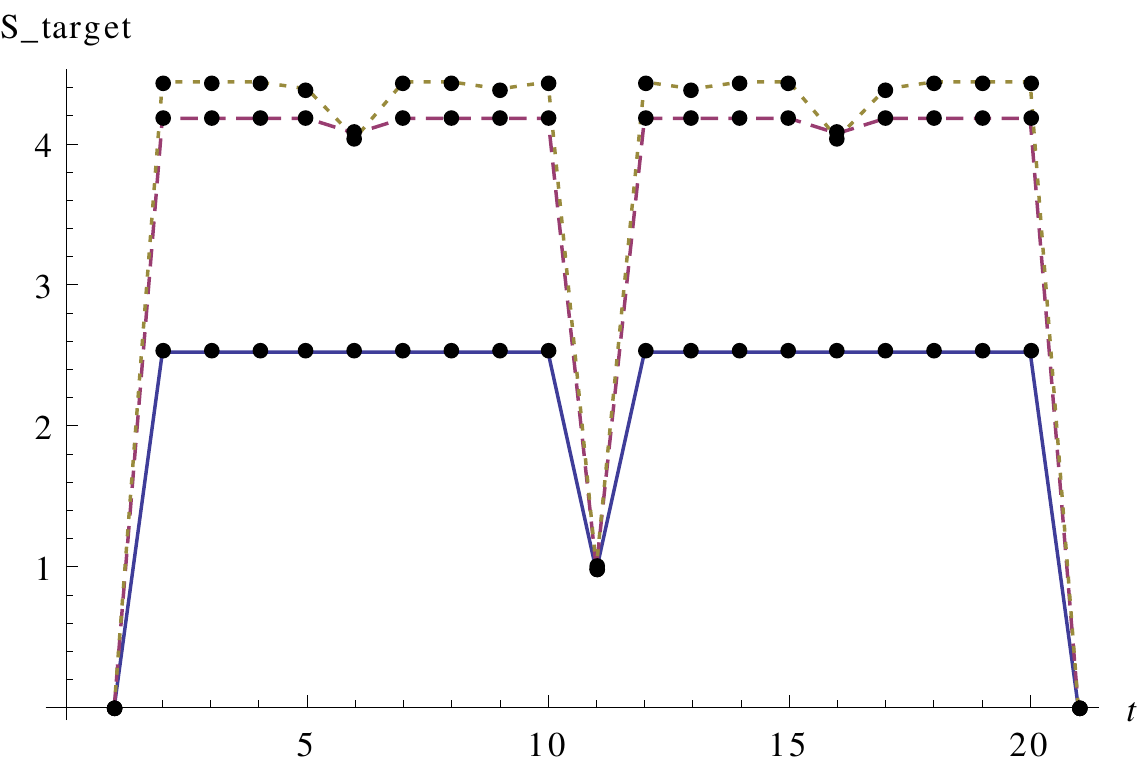} \hfill \includegraphics[width=0.3\textwidth]{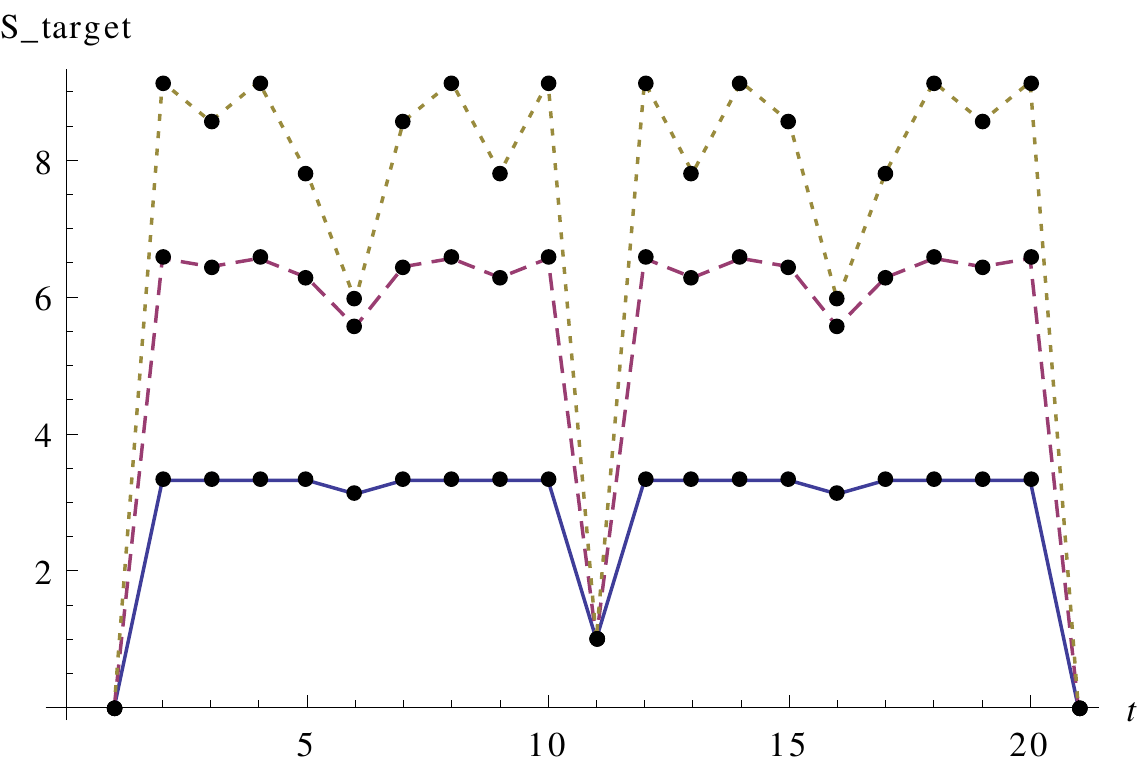}\\
\includegraphics[width=0.3\textwidth]{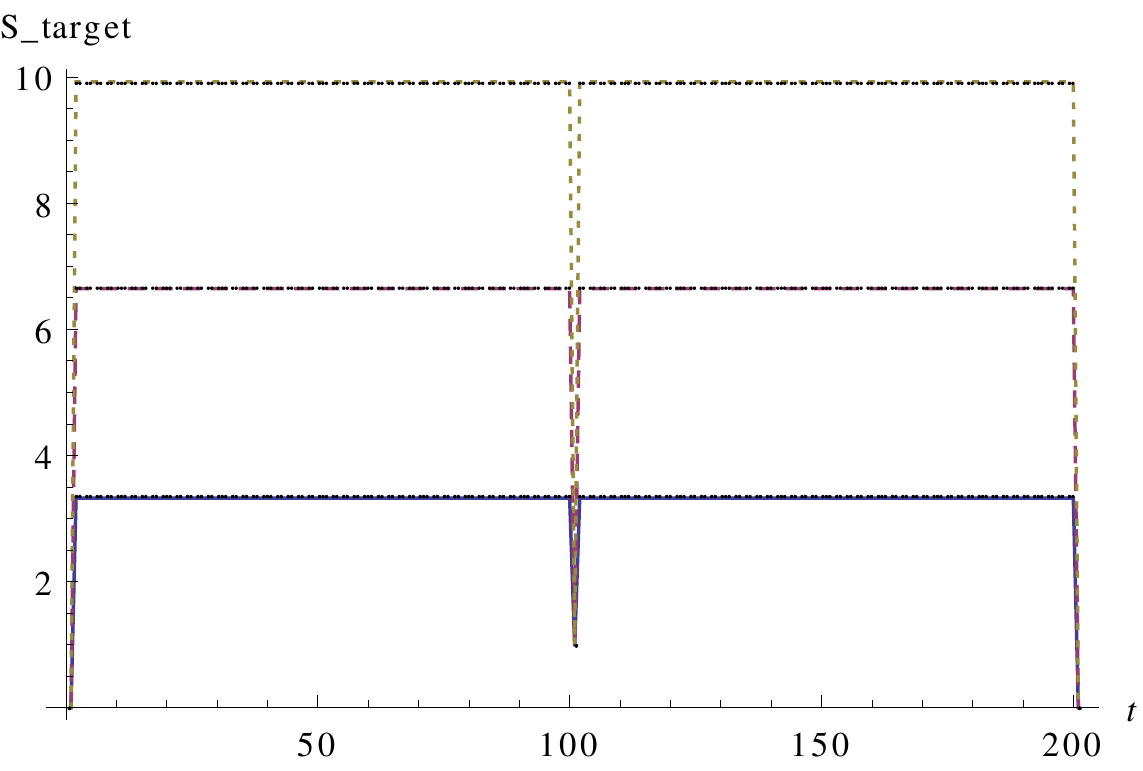} \hfill \includegraphics[width=0.3\textwidth]{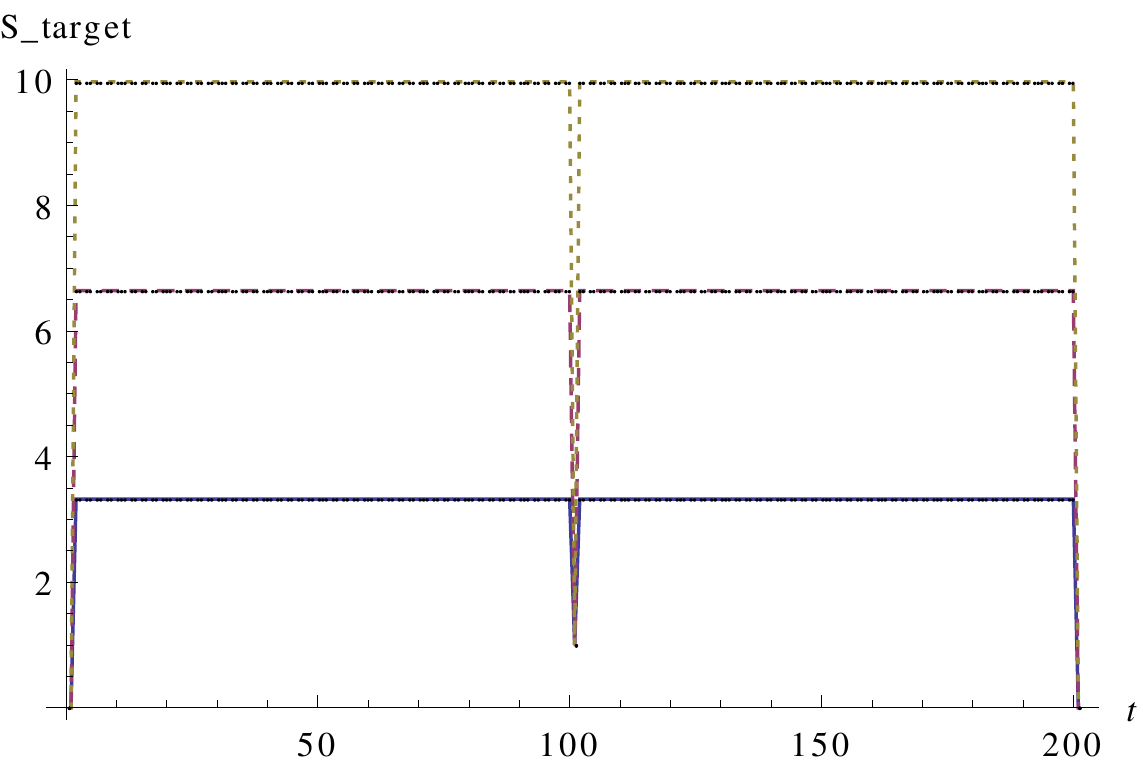} \hfill \includegraphics[width=0.3\textwidth]{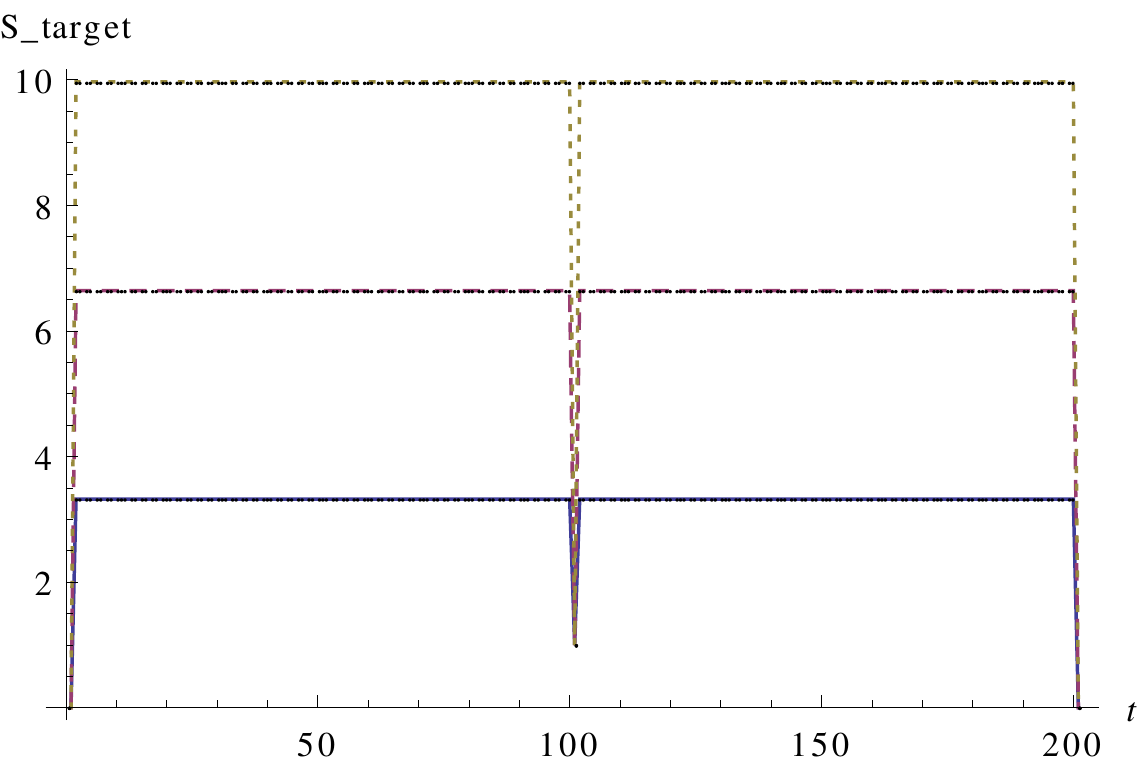}\\
\includegraphics[width=0.3\textwidth]{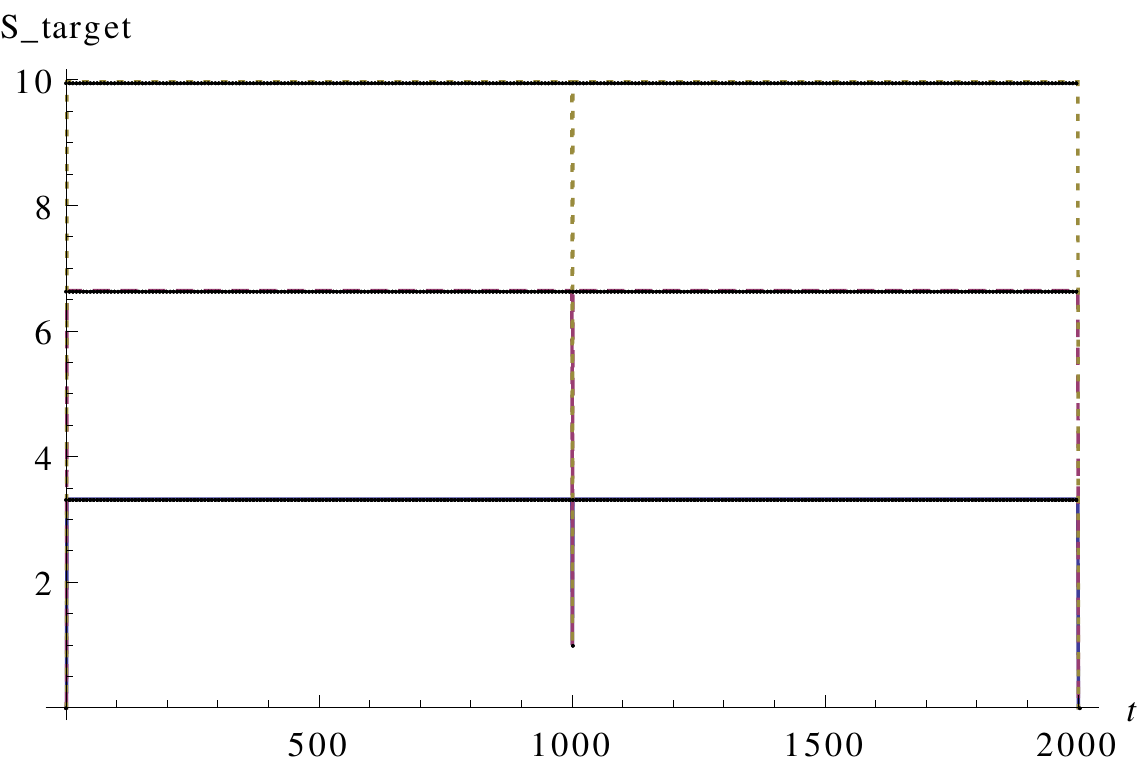} \hfill \includegraphics[width=0.3\textwidth]{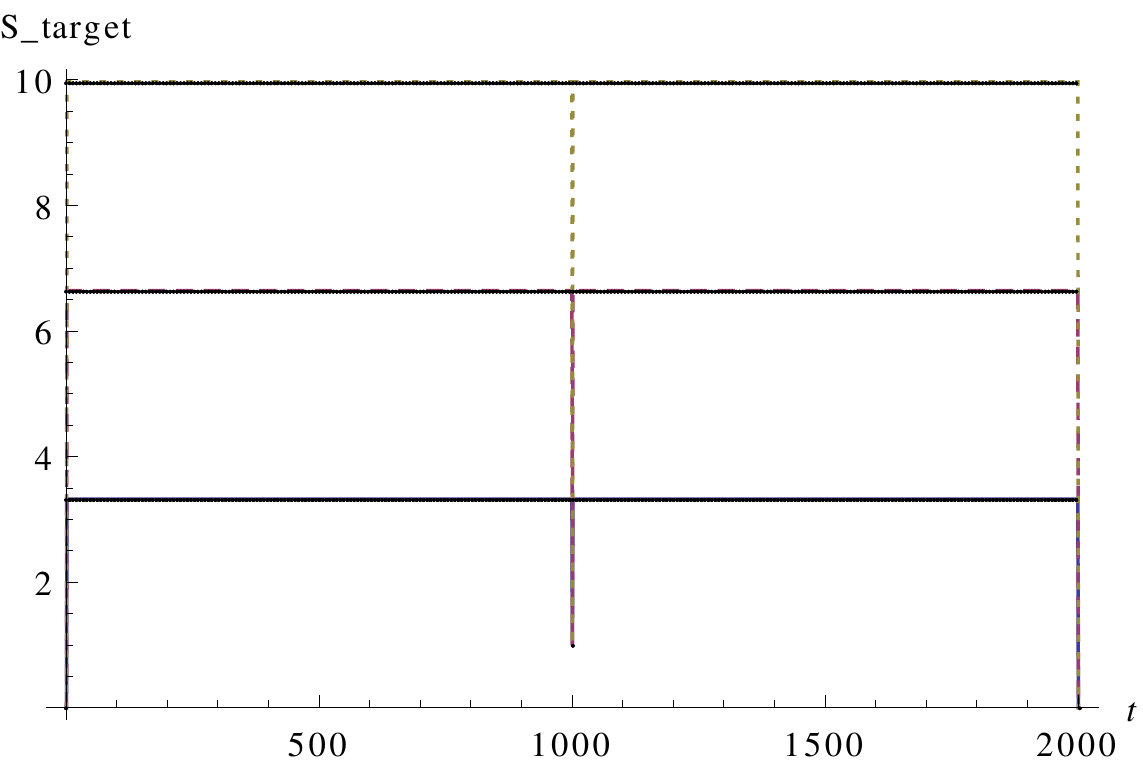} \hfill \includegraphics[width=0.3\textwidth]{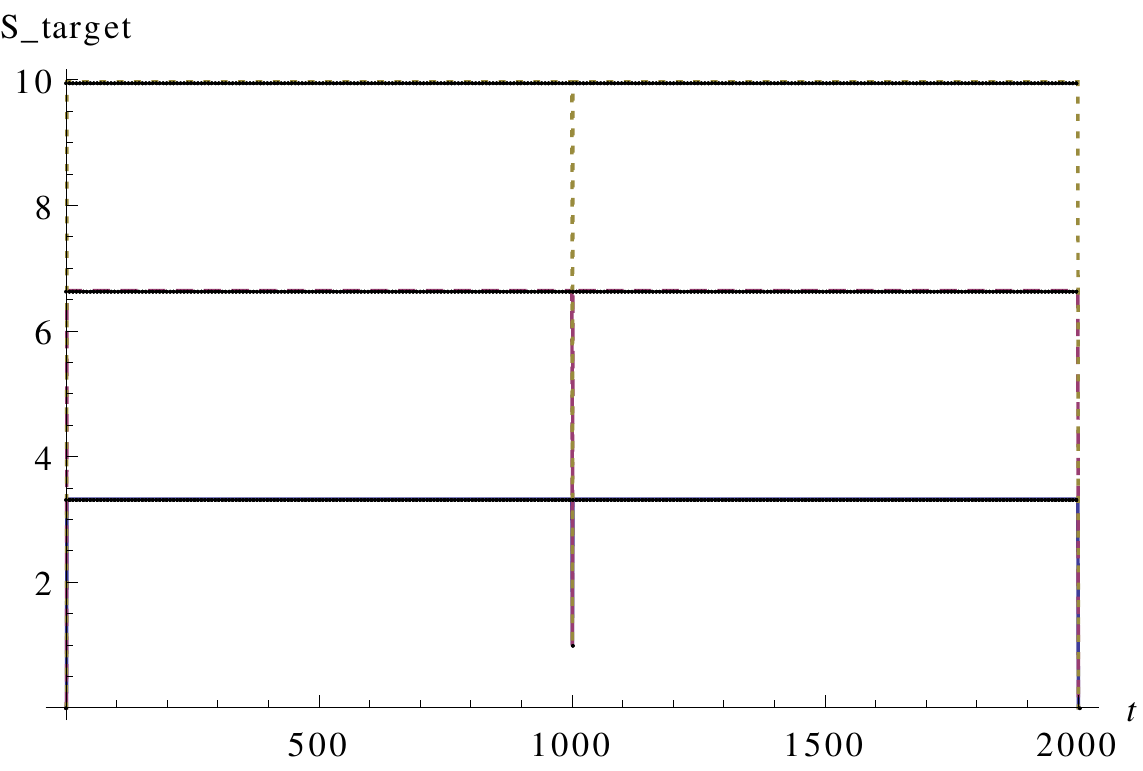}
\caption{The time development of the entanglement entropy for a superposition of the control ring drawn from a marker distribution around an average magnetization $\mu = 0.05$ (left column), $\mu = 0.1$ (middle column) and $\mu = 0.5$ (right column) for rings of length $N=10$ (first row), $N=100$ (second row) and $N=1000$ (third row). In each plot, the control ring consists of $M=10$ (solid line), $M=100$ (dashed line) and $M=1000$ (dotted line) randomly chosen states in superposition.}
\label{fig:entropyformu}
\end{figure}

We can use this relation to the classical Kac ring to estimate the evolution of entanglement entropy for more general and also more physical control ring configurations. Now, for the superposition of the control ring, we randomly choose $M$ rings from a distribution which is centered around an average density $\mu$ of $\ket{1}$'s. We have examined the entanglement entropy depending on $N$, $M$ and $\mu$ (see figures \ref{fig:entropyformu}). We observe that for long enough rings, the entanglement entropy is determined by the number of states in the control ring, which is the number of classically available configurations. Assuming that the diagonal entries of the reduced density matrix of the target ring are more or less equal for the maximally entangled state, we find that the maximum of the entanglement entropy should be about
\begin{equation}
\label{eq:Sapprox}
 S_{\text{target}}^{M} = M \frac1M \log_2 \frac1M \;.
\end{equation}
For the examples considered, this is $S_{\text{target}}^{10} \approx 3.3, S_{\text{target}}^{100} \approx 6.6$ and $S_{\text{target}}^{1000} \approx 10.0$, which agrees with the second and third lines in Fig.~\ref{fig:entropyformu}. However, if the ring is too short, the number of possible configurations $\binom{N}{\mu N}$ can be smaller than the number $M$ of rings in the sample and \eqref{eq:Sapprox} overestimates the entropy. This happens in the first line in the figures where the entropy also depends on $\mu$ and $N$. For small values of $\mu$, the entropy saturates also for larger $M$ (dashed and dotted lines in the top left plot of Fig.~\ref{fig:entropyformu}). For larger values of $\mu$, there are comparably many possibilities to distribute the markers over the different sites and the number of possible classical states is comparably high. It becomes evident that the capacity of the quantum Kac ring to store entanglement really depends on how many states contribute to the superposition of the control ring rather than the size of the system.

\subsection{Magnetization}

Another interesting observable is the magnetization in the target ring $\langle \hat{S}_z \rangle$. This is the quantum analogue of the grayness $\Delta(t)$ in the classical ring. There is a crucial difference between the two observables. While the entanglement entropy is derived from the density matrix, the magnetization is derived from the state itself. Therefore, it will be more sensitive to phase changes, which drop out when computing the density matrix.

For the initial state of the target ring $\ket{00 \dots 0}$, which we prepare, the magnetization is $\langle \widehat{M} \rangle = \langle \hat{S}_z (0) \rangle =  N$, because the $\ket{0}$-qubit encodes a spin up. Obviously, the subsequent evolution of the magnetization depends strongly on the configurations of the control ring, which one is allowing for. In the case of the maximal superposition of the control ring, which we have been using so far, every combinatorially possible combination of spin flips occurs exactly once in one superposition
\begin{eqnarray}
\cnot \left(\biket{2}{0 \\1}+ \biket{2}{1 \\1}\right) & = & \biket{2}{0 \\1}+ \biket{2}{1 \\0} \\
\cnot \left(\biket{2}{0 \\0}+ \biket{2}{1 \\0}\right) & = & \biket{2}{0 \\0}+ \biket{2}{1 \\1} \;.
\end{eqnarray}
This means, that for the average of all the superpositions, $\langle \hat{S}_z (1) \rangle=0$ after one time step. It remains naught until the recurrence time, $\langle \hat{S}_z (t) \rangle = 0 \; \forall 0<t<2N$, after which it obviously returns to $\langle \hat{S}_z (2N) \rangle = N$.

\subsection{Thermalization}

Since the energy levels of our system are highly degenerate, at least without an external field, we need to specify what we mean with a thermal state. First, a thermal state should be macroscopically stationary, that means that thermodynamic quantities like the entanglement entropy and the magnetization should not change much. Secondly, the state occupancy should strive towards an equal distribution over all possible states for long times. In our case, this would mean that all the diagonal elements of the density matrix have equal probability, namely $1/2^N$. This is the state of maximal entropy. According to the second law, this should be the final state of the system. 

The intriguing observation is now, that for the development of the quantum Kac ring, this state is achieved instantaneously.  The system ``jumps'' into the ``thermodynamically final'' state. After that, the entropy drops slightly and oscillates around a value close to the maximum on scales smaller than $N$, at which point it suddenly drops to a minimum but non-zero value of one. In the second half of the evolution, the first half is repeated in reverse due to the microscopic time reversal invariance. Note that at any intermediate state, comparing the increase or decrease of entropy towards both the preceding as well as the following state, does not allow to decide the direction of time. Every possible entropy change, which would occur after reversing the evolution always also occurs in the forward direction.

If we start with a state, which is separable between target and control ring, we know that after minimally $2N$ steps we arrive again at a product state, whose entanglement entropy vanishes. This means that thermalization can occur only for a time scale smaller than the system size and much smaller than the dimension of the Hilbert space of this system.


\section{Typical vs. average behavior of measurements in the quantum ring}
\label{sec:typicality}

Since we perceive the control ring, which is prepared in a superposition as the environment, it is interesting to examine the typical outcome of an experiment and compare it to the classical case. 

The crucial difference between the classical and the quantum system is that a measurement on a classical system with a specific marker configuration will always yield the same result, while in the quantum model a specific result will be obtained with its corresponding probability. The quantum ring is a parallel computing device, in which several classical configurations are produced in superposition. A typical measurement in the classical case is thus performed by averaging several measurements on an ensemble of rings with random distribution of markers. For the quantum case the typical result is already produced by performing a sequence of measurement on equally prepared quantum systems.

Let us first consider the case, where the probabilities of finding a value of $\ket{1}$, i.e. a marker or a spin up, is the same on each site of the control ring. These probabilities are then the average magnetization $\mu$ and they are independent of each other. In that case, the target ring consists of a superposition of all the elements of a typical ensemble of classical Kac rings. Successive measurements on the quantum ring will then yield the same result as successive measurements on randomly chosen different elements of an ensemble of classical rings. 

Remarkably, also the variance and in fact all of the higher moments of the result of the quantum measurement are the same as in the classical case. This is so, because the distributions are, indeed, identical. To see this, we examine the contribution of each term in the superposition of the target ring to the magnetization. We label each such term with $T^{(i)}(t)$ such that the state of the target ring is
\begin{equation}
\ket{T (t)} = \ket{T^{(1)}(t)} + \ket{T^{(2)}(t)} + \dots \;.
\end{equation}
The magnetization $\langle \widehat{M} \rangle (t) = \langle T (t) | \widehat{M} | T(t) \rangle$ then breakes apart into the contribution
\begin{equation}
\langle \widehat{M}^{(i)} \rangle (t)= \bra{T^{(i)}(t)} \widehat{M} \ket{T^{(i)}(t)} = \frac1N \bra{T^{(i)}(t)}\bigoplus_{j=1}^N \hat{S}_z^{(j)} \ket{T^{(i)}(t)}
\end{equation}
of each term. This measures the classical observable $\Delta(t)$ for a specific target ring $\ket{T^{(i)}}$ in the superposition.

Measurement of the magnetization $\bra{T} \widehat{M} \ket{T}$ projects the target ring onto one of its states, because the possible states of the target ring are orthogonal eigenstates of $\widehat{M}$, such that $\braket{T^{(i)}}{T^{(j)}} = 0$ if $i \neq j$. Note that during the time evolution, several terms in the superposition of the target ring can become identical and the norm of the $T^{(i)}(t)$ changes with time. We denote the number of rings which thus become identical by $a_i(t)$ such that $\braket{T^{(i)}(t)}{T^{(i)}(t)}  = a_i^2(t)$. In this notation, every constituent of the target ring contributes the magnetization 
\begin{equation}
 \langle \widehat{M}^{(i)} \rangle (t)= m_i(t) w_i (t) \;,
\end{equation}
where $w_i(t)= a_i^2(t)/A^2$ is the weight of each term in the target ring and $m_i(t)$ its magnetization. 
The expectation value of the total magnetization 
\begin{equation}
\langle \widehat{M} \rangle (t) =  \sum_i \langle \widehat{M}^i \rangle = \sum_i m_i(t) w_i(t)
\end{equation}
yields the average over the results of measurements of different classical systems in a randomly chosen sample.
Note, however, that generically some of the $m_i(t)$ are the same because different configurations of target rings with the same number of spins up or down have the same magnetization. This is just as in the classical model where the grayness $\Delta(t)$ does not depend on the precise microscopic order of the balls. 
The probability for measuring a specific value $M$ for the magnetization is then $\sum_{i | m_i = M } w_i$, which scales with the number of rings with magnetization $M$ in the superposition. This corresponds to the number of classical Kac rings with a specific $\Delta(t)$. The distribution and thus also all the moments for the magnetization are the same as for $\Delta(t)$ if the superposition of the control ring has been prepared such that drawing a part of the superposition at random generates a distribution of spins pointing up and acting as markers equal to the distribution of markers in the classical case, usually a binomial distribution.
Note that just as in the classical case, the expectation value for the magnetization might not be typical for any of the outcomes of a specific measurement.

We can generalize this finding to a slightly more realistic situation: We now fix the magnetization in the control ring, i.e. the precise number of markers. Can we still use the classical results to estimate the result of an experiment?

Let $N$ be the number of sites and $n$ be the number of markers. That means, there are $i=N-n$ non-marker sites in the control ring. We are interested in the probability of finding $j$ markers on (the first) $t$ consecutive sites for a chain of length $N$ with $n$ markers in total (cf. discussion in section \ref{sec:micdyn}, in particular equation \eqref{eq:markeraverage}). We denote this probability with $p^{N,n}_j(t)$. This probability is the number of occurrences of precisely $j$ markers on the first $t$ sites divided by all the possibilities to distribute the markers over all the sites arbitrarily $\binom{N}{n}$. 

If there are as many markers as sites $n=N$, there is only one possibility to distribute them. if there are some non-marked sites, we have to distribute $j$ markers over the first $t$ sites and the remaining $t-j$ markers over the remaining $N-t$ sites. Note, that this means that for a given $t$, there are at most $N-n+1$ non-vanishing probabilities. We find as the result
\begin{equation}
p^{N,n}_j (t) = \frac{\binom{t}{j} \binom{N-t}{n-j}}{\binom{N}{n}} \;.
\end{equation}

Now the next question is if there is a connection to the binomially distributed markers in the thermodynamic limit $N \to \infty$ while $\mu = \frac{n}{N} = \const$
\begin{equation}
p^{N,n}_j(t) \stackrel{N \to \infty}{\longrightarrow} p_j(t) = \binom{t}{j} \mu^j (1-\mu)^{t-j} \;.
\end{equation}

To take this limit, we write out the binomial coefficients 
\begin{equation}
\frac{\binom{N-t}{\mu N -j}}{\binom{N}{\mu N}} = \frac{(N-t)! (\mu N)! (N-\mu N)!}{(\mu N -j)! (N-t- \mu N +j)! N!} 
\end{equation}
and use Stirling's approximation $x! \approx \left( \frac{x}{\ee} \right)^x \sqrt{2 \pi x}$. We see immediately that the powers of $\ee$ cancel. In the limit of large $N$, the exponentials go to
\begin{equation}
\lim_{N\to \infty} \frac{N^{N-t} (\mu N)^{\mu N} (N(1-\mu))^{N(1-\mu)}}{(\mu N)^{\mu N - j} (N(1-\mu))^{N(1-\mu)- (t-j)}} = \mu^j (1- \mu)^{t-j}
\end{equation}
which is precisely the binomial distribution. Hence, in the thermodynamic limit, we can allow for another class of states in the control ring, where the total magnetization is fixed rather than all the individual magnetizations at each site. This is reminiscent of the equivalence between the micro-canonical and the canonical ensemble, where the average energy is fixed macroscopically instead of fixing the individual energies.

\section{Conclusions}
\label{sec:conclusions}

Classical and quantum many body systems are microscopically time reversal invariant. This symmetry would have to be broken for thermalization to occur. The Kac ring is a simple toy model in which it becomes very clear that neglecting correlations in the target ring amounts to making a Boltzmann approximation. This is valid only up to a certain time which we can here derive explicitly. Beyond this approximation, we can also examine the anti-Boltzmann behavior. It is easily seen that the recurrence time is much shorter than the dimension of the Hilbert space. 

Since the quantum Kac ring can support superpositions, it computes a number of classical evolutions in parallel. This gives a difference in how the result of a measurement is determined between a classical and a quantum system. While in a classical system, the result is an average over measurements on systems drawn from an ensemble of systems with different control ring configurations, for the quantum system, measuring equally prepared typical systems will already reproduce the classical result. 

The model we present is meant as a pedagogical example, in which we can explicitely study a realization of the molecular chaos assumption leading to a quantum Boltzmann equation and its range of validity as well as the question of typicality of the result of measurements in a quantum many body system

\section*{Acknowledgments}

This work was supported through SFB 1073 of the Deutsche Forschungsgemeinschaft (DFG).

\bibliographystyle{elsarticle-num}
\bibliography{qKac}

\end{document}